
\magnification=1200
\baselineskip=18pt
\tolerance=100000
\overfullrule-0pt
 \def\bull{\vrule height .9ex width .8ex depth 0pt}
\centerline{\bf REPRESENTATIONS OF CLIFFORD ALGEBRAS}
\centerline{\bf  AND ITS APPLICATIONS}

\bigskip

\centerline{by}

\bigskip

\centerline{Susumu Okubo}
\centerline{Department of Physics and Astronomy}
\centerline{University of Rochester}
\centerline{Rochester, NY 14627, U.S.A.}

\vskip 1.7 truein

\noindent {\bf \underbar{Abstract}}

\medskip

A real representation theory of real Clifford algebra has been studied in
further detail, especially in connection with Fierz identities.  As its
application, we have constructed real octonion algebras as well as
related octonionic triple system in terms of 8-component spinors
associated with the Clifford algebras $C(0,7)$ and $C(4,3)$.

\vskip 1.8 truein

\noindent AMS 15A66.,15A69.

\vfill\eject

\noindent {\bf 1. \underbar{Introduction}}

\medskip

This paper is dedicated to the memory of the late Professor Eduardo R.
Caianiello.  It may be remarked that both Prof. Caianiello and I were
Ph.D. graduates of the University of Rochester with Ph.D. degrees awarded
respectively in 1950 and 1958.  However, my professional contact with him
began in 1959, when he kindly invited me to the University of Napoli as a
research associate.
  Although I stayed there only one year, I was
always enchanted by his personal warmness and charm.  Professionally, his
relaxed but yet great curiosity toward sciences in general
 greatly influenced my subsequent career in
physics.

This paper is about the applications of Clifford algebras, in which Prof.
Caianiello had also worked in his earlier researches ([1], [2], [3],
[4]).  The $N$-dimensional real Clifford algebra $C(p,q)$ with $N = p+q$
is an associative algebra generated by Dirac matrices $\gamma_\mu \ (\mu
= 1,2,\dots , N)$ satisfying
$$\gamma_\mu \gamma_\nu + \gamma_\nu \gamma_\mu = 2 \eta_{\mu \nu} E
\eqno(1.1)$$
where $\eta_{\mu \nu} \ (\mu ,\nu =1,2,\dots ,N)$ are constants given by
$$\eta_{\mu\nu} = \cases{0\ , &if $\mu \not= \nu$\cr
\noalign{\vskip 5pt}%
1\ , &if $\mu = \nu = 1,2, \dots , p$\cr
\noalign{\vskip 5pt}%
-1\ , &if $\mu = \nu = p+1 ,p+2,\dots ,N$ \quad . \cr}\eqno(1.2)$$
In Eq. (1.1), $E$ stands for the unit matrix.  The representation theory
of the complex Clifford algebra is well-known ([5], [6], [7]).  The
representation module is fully reducible, and the dimension $d$ of the
irreducible representation space (hereafter referred to as IRS) is given by
$$d = 2^n$$
for both cases of $N=2n$ and $2n +1$ irrespective of $p$ and $q$.
Moreover, for $N=2n$, the IRS is unique, while we have two inequivalent
 IRS for $N=2n +1$ which are related to each
other by $\widetilde \gamma_\mu = - \gamma_\mu$.

However, the real representation theory of the real Clifford algebra
$C(p,q)$ is more involved.  We may approach the problem in the following
two ways.  We may start from the classification theory of $C(p,q)$ by
Proteous [8], as has been done by Hile and Lounesto [9].  The second method
is to utilize a theorem of Frobenius [10] on real division algebra as in
ref. [11].  The advantage of the latter method which we will use in
this paper is its relevance to physical applications when we require
properties of the charge conjugation matrix $C$ as well as of the Fierz
transformation.  Because of this, we will first briefly sketch some of
 the relevant facts needed in this paper.

Let $\gamma_\mu$ be the $d \times d$ real irreducible representation
matrices of the real Clifford algebra $C(p,q)$ with
$$N = p + q \quad . \eqno(1.3)$$
Suppose that we have a real $d \times d$ matrix $S$ satisfying
$$[ S, \gamma_\mu ] = 0 \qquad (\mu = 1,2,\dots , N) \quad . \eqno(1.4)$$
Then, the standard reasoning based upon Schur's lemma requires $S$ to be
invertible, unless it is identically zero.  Hence, a set consisting of
all $d \times d$ matrices $S$ satisfying Eq. (1.4) defines a real
associative division algebra.  As the consequence of the Frobenius
theorem, we will have the following three cases:

\medskip

\noindent {\bf (I) \underbar{Normal Representation}}

\medskip

We must have
$$S = a E \eqno(1.5)$$
for some real constants $a$, where $E$ is the $d \times d$ unit matrix.

\medskip

\noindent {\bf (II) \underbar{Almost Complex Representation}}

\medskip

The general solution of Eq. (1.4) is given by
$$S = aE + bJ \eqno(1.6)$$
for some real constants $a$
 and $b$, where real $d\times d$ matrix $J$
 satisfies
$$\eqalignno{&J^2 = - E \quad , &(1.7a)\cr
&[J , \gamma_\mu ] = 0  \quad .&(1.7b)\cr}$$

\vfill\eject

\medskip

\noindent {\bf (III) \underbar{Quaternionic Representation}}

\medskip

There exist three real $d \times d$ matrices $E_j \ (j = 1,2,3)$ which
commute with $\gamma_\mu$'s and satisfy the quaternionic relation:
$$\eqalignno{&[ \gamma_\mu , E_j ] = 0 &(1.8a)\cr
&E_j E_k = - \delta_{jk} E + \sum^3_{\ell =1}
\epsilon_{jk \ell} E_\ell \quad , \quad (j,k = 1,2,3) \quad , &(1.8b)\cr}$$
where $\epsilon_{jk \ell}$ is the totally antisymmetric Levi-Civita
symbol in three dimensional space.  We now must have
$$S = a_0 E + \sum^3_{j=1} a_j E_j \eqno(1.9)$$
for real constants $a_0$, $a_1$, $a_2$, and $a_3$.

Since any real representation of $C(p,q)$ can be shown to be fully
reducible, we will consider only the case of irreducible representations
in the following unless stated otherwise.  We have now proved elsewhere
[11] the following results:

\medskip

\noindent {\bf (I) \underbar{Normal Representation}}

\medskip

This case is possible if and only if we have
$$p - q = 0 , 1, 2 \ \ ({\rm mod}\ 8) \eqno(1.10)$$
with dimension
$$d = 2^n \eqno(1.11)$$
for
$$N = p + q = 2n \ \ {\rm or}\ \ 2n + 1 \quad . \eqno(1.12)$$
Moreover, it is unique for $N = 2n$, while we have two inequivalent IRS
for the case of $N = 2n +1$, which are related to each other by
$\widetilde \gamma_\mu = - \gamma_\mu$.  Also, we must have
$$\gamma_1 \gamma_2 \dots \gamma_N = \pm E \eqno(1.13)$$
for $N = 2n +1$, but not for $N=2n$.

\medskip

\noindent {\bf (II) \underbar{Almost Complex Representation}}

\medskip

The almost complex representation is realizable if and only if we have
$$p - q = 3\ {\rm or}\ 7\ \ ({\rm mod}
\  8) \eqno(1.14)$$
 so that $N = p+q$ is always odd.  The IRS is unique with dimension
 $$d = 2^{n+1} \quad , \eqno(1.15)$$
 while the $d \times d$ matrix $J$
  satisfying  Eqs. (1.7) is given by
 $$J = \pm \gamma_1 \gamma_2 \dots \gamma_N \quad . \eqno(1.16)$$
 Moreover, there exists another $d \times d$ matrix $D$ satisfying
 $$\eqalignno{&D \gamma_\mu + \gamma_\mu  D = 0 &(1.17a)\cr
 &D^2 = (-1)^{{1 \over 4}\ (p-q +1)} \quad . &(1.17b)\cr}$$

 \medskip

\noindent {\bf (III) \underbar{Quaternionic Representation}}

\medskip

This case can occur if and only if we have
$$p-q = 4,5, 6 \ \ ({\rm mod} \ 8) \eqno(1.18)$$
with the dimension
$$d = 2^{n+1} \quad . \eqno(1.19)$$
The IRS is unique for $N =$ even, while there exists two inequivalent IRS
related to each other by $\widetilde \gamma_\mu = - \gamma_\mu$ for the
case $N =$ odd.  Moreover, for the latter case, we have also the validity
of Eq. (1.13).

Finally, let us comment upon the charge conjugation matrix $C$.  In [11],
we have shown the existence of real charge conjugation matrix $C$
satisfying the following properties:

\vfill\eject

\medskip

\noindent {\bf \underbar{Case 1}}

\medskip

$$C \gamma_\mu C^{-1} = - \gamma_\mu^T \eqno(1.20a)$$

\medskip

\noindent {\bf \underbar{Case 2}}

\medskip

$$C \gamma_\mu C^{-1} = + \gamma_\mu^T \eqno(1.20b)$$
where $Q^T$ stands for the transpose of a matrix $Q$.  Actually, this
distinction of cases 1 and 2 is somewhat arbitrary for the case $N =$
even, or for the almost complex representation, since we can change the
signs in the right sides of Eqs. (1.20) by replacing $C$ by $CQ$ with $Q
= \gamma_1 \gamma_2 \dots \gamma_N$ for $N =$ even, and by $CD$ for the
almost complex representation.  However, for both normal and quaternionic
representations for the $N =$ odd case, such a transformation is not
possible and we must maintain the distinction.  Keeping this fact in
mind, then we find that the exceptional case 2 is possible only if we
have $p -q = 1$ or 5 (mod 8) with $N = 4 \ell +1$ for some integer
$\ell$. We assign all other cases to the case 1 in what follows.

The charge conjugation matrix $C$ obeys the relation
$$C^T = \eta C  \eqno(1.21)$$
for $\eta = \pm 1$.  The parity $\eta$ can be given by
$$\eta = (-1)^\ell \quad , \eqno(1.22a)$$
when we have
$$\eqalign{
&{\rm (i)}\quad  p-q = 0 \ {\rm or}\  2\
 {\rm (mod} \  8) \ {\rm with}\  N=4\ell\cr
&{\rm (ii)}\quad  p-q =4\ {\rm or}\  6 \ ({\rm mod}\  8)\
 {\rm with}\  N=4\ell + 2\cr
&{\rm (iii)}\quad  p-q =1\  ({\rm mod}\  8)
\  {\rm with}\  N =4\ell +1\cr
&{\rm (iv)}\quad  p-q =5\  ({\rm mod}\  8)\
 {\rm with}\  N= 4\ell +3 \quad ,\cr}\eqno(1.22b)$$
 while we find
 $$\eta = (-1)^{\ell +1} \eqno(1.23a)$$
 for the cases of
 $$\eqalign{
&{\rm (v)}\quad  p-q = 0\ {\rm or}\  2\
 {\rm (mod}\  8) \ ,\  N=4\ell +2\cr
&{\rm (vi)}\quad  p-q =4\ {\rm or}\  6 \ ({\rm mod}\  8)\
  N=4\ell\cr
&{\rm (vii)}\quad  p-q =1,3, \  {\rm or}\ 7
\ ({\rm mod}\  8) \ ,\
N =4\ell +3\cr
&{\rm (viii)}\quad  p-q =5\  ({\rm mod}\  8)\ ,\
 \  N= 4\ell +1 \quad ,\cr}\eqno(1.23b)$$
 while the sign of $\eta$ is undeterminable even in principle for $p-q
 =3$ or 7 (mod 8) with $N = 4\ell +1$ since we can change the sign of
 $\eta$ at will by using $CJ$ instead of $C$.  However, the parity of
 $CD$ can then be determined as in [11].

 For quaternionic representation, we have
 $$CE_j C^{-1} = - E^T_j \qquad (j = 1,2,3)
  \quad .\eqno(1.24)$$

 \medskip

 \noindent {\bf \underbar{Remark 1.1}}

 \medskip

 The charge conjugation matrix $C$ in this paper corresponds to its
 inverse $C^{-1}$ of the ref. [11].

 \medskip

 \noindent {\bf \underbar{Remark 1.2}}

 \medskip

 There exist some isomorphisms among real Clifford algebras.  Consider
 $C(p,q)$ and $C(p^\prime , q^\prime)$ with $p^\prime + q^\prime = p + q
 (\equiv N)$.  If we have either
 $$\eqalignno{&{\rm (i)}\quad p^\prime - q^\prime = p - q \ ({\rm mod}\
 8)\qquad\qquad\qquad\qquad &(1.25a)\cr
 {\rm or}\qquad\qquad &\cr
 &{\rm (ii)}\quad p^\prime - q^\prime + p - q = 2\ ({\rm mod}\ 8) \quad ,
 \qquad\qquad\qquad\qquad&(1.25b)\cr}$$
 then $C(p^\prime , q^\prime)$ is isomorphic
  to $C(p,q)$.  See refs. [8] and also [12].  This fact is
  consistent with our results stated above.

  \medskip

  \noindent {\bf \underbar{Remark 1.3}}

  \medskip

  As we noted, the almost complex representation is possible only for the
  case of $N= 2n+1$ being odd.  It is intimately related to the normal
  representation of $N+1$ dimensional Clifford algebra.  Suppose that
  $p-q = 3$ (mod 8).  Then, setting $D = \gamma_{N+1}$, we see that
  $\gamma_1$, $\gamma_2$, $\dots$, $\gamma_N$, and $\gamma_{N+1}$
   form a real normal
  representation $C(p,q+1)$ of the even-dimensional Clifford algebra.
  For the other case of $p-q = 7$ (mod 8), we identify $D= \gamma_0$.
  Then, $\gamma_0$, $\gamma_1$, $\dots$, $\gamma_N$ will define the
  normal IRS of $C(p +1, q)$.

  \medskip

  \noindent {\bf \underbar{Remark 1.4}}

  \medskip

  If we choose a suitable basis, then we can assume the validity of
  $$\gamma_\mu^T = \cases{\gamma_\mu \ ,\  & $\mu = 1,2, \dots, p$\cr
  \noalign{\vskip 5pt}%
  -\gamma_\mu \ ,\ &  $\mu = p + 1 , \dots , N$ \cr}\eqno(1.26a)$$
  as well as
  $$E_j^T = - E_j \quad , \quad J^T = - J \quad , \quad D^T = (-1)^{{1
  \over 4}\ (p -q+1)} D \quad . \eqno(1.26b)$$
  See ref. [11] for details.

  \medskip

  \noindent {\bf \underbar{Remark 1.5}}

  \medskip

  The Clifford algebra $C(p,q)$ is intimately related to the non-compact
  orthogonal group SO($p,q$).  Setting
  $$J_{\mu \nu} = {1 \over 4}\ [ \gamma_\mu , \gamma_\nu ] \quad ,
  \eqno(1.27)$$
  it defines the Lie algebra so($p,q$):
  $$\left[ J_{\mu \nu} ,J_{\alpha \beta} \right] = \eta_{\nu \alpha}
  J_{\mu \beta} - \eta_{\nu \alpha} J_{\alpha \mu} + \eta_{\nu \beta}
  J_{\alpha \mu} - \eta_{\mu \beta} J_{\alpha \nu} \quad . \eqno(1.28)$$
  Let
  $$\omega^{\mu \nu} = - \omega^{\nu \mu} \qquad (\mu , \nu = 1,2,
  \dots , N) \eqno(1.29)$$
  be some real constants and set
  $$U = \exp \left\{
  {1 \over 2}\ \sum^N_{\mu , \nu =1} \omega^{\mu \nu} J_{\mu \nu}
  \right\} \quad . \eqno(1.30)$$
  Then, it satisfies the ortho-symplectic condition
  $$U^T \ C\ U = C \eqno(1.31)$$
  when we utilize Eqs. (1.20).  Such a $U$ defines the spinor
  representation of the spin group which is a covering of the SO($p,q$).
  $\bull$

  Our paper is organized as follows.  In the next section, we will study
 some additional property of the quaternionic representation with some
 comments about ref. [3]
 of the papers by Caianiello.  In section 3, we will study orthogonality
 relations, as well as the resulting Fierz identities, and find that the
 so-called Burnside theorem will fail for both almost complex and
 quaternionic representations.  In section 4, we will construct
 real and complex octonion algebras out of real and complex spinors for
 the Lie algebra so(7) as an application of the Fierz transformation,
 while we will make some comment on the dimension of real Hurwitz
 algebras in section 5.

 \medskip

 \noindent {\bf 2. \underbar{Quaternionic Representation}}

 \medskip

 We will now explain the relationship between our quaternionic IRS and
 the standard matrix realization
  ([8], [9]) over the real quaternionic division
 algebra.

 Let $e_j \ (j = 1,2,3)$ and the unit $e_0$ be the basis vectors of real
 quaternionic division algebra, satisfying
 $$e_j e_k = -\delta_{jk} e_0 + \sum^3_{\ell =1} \epsilon_{jk \ell}
 e_\ell \quad . \eqno(2.1)$$
 Let $\rho (e_j)$ be real
  irreducible matrix realizations of $e_j$.  When we
 note
 $$\rho (e_j) \rho (e_k) + \rho (e_k) \rho (e_j) = -2 \delta_{jk} \rho
 (e_0) \quad , \eqno(2.2)$$
 it corresponds to the 4-dimensional
 quaternionic representation of the real Clifford algebra
 $C(0,3)$ in view of Eqs. (1.18) and (1.19).  Hence, there exist
 another $4 \times 4$  real
 quaternionic matrices $\hat E_j \ (j = 1,2,3)$
 satisfying
 $$\eqalignno{&\big[ \rho (e_j) , \hat E_k \big] = 0&(2.3a)\cr
 &\hat E_j \hat E_k = -\delta_{jk} \hat E_0 + \sum^3_{\ell =1}
 \epsilon_{jk \ell} \hat E_\ell &(2.3b)\cr}$$
 where $\hat E_0 = \rho (e_0)$
  stands for the $4\times 4$ unit matrix.  In terms of $2
 \times 2$ Pauli matrices $\sigma_1$, $\sigma_2$, and $\sigma_3$ as well
 as the $2 \times 2$ unit matrix $E_0$, we may identify
 $$\eqalignno{&\rho (e_1) = i \sigma_2 \otimes E_0\quad  , \quad
 \rho (e_2) = \sigma_1 \otimes i \sigma_2 \quad  ,
\quad \rho (e_3) = \sigma_3 \otimes i \sigma_2 \quad  ,
 &(2.4a)\cr
 &\hat E_1 = E_0 \otimes i \sigma_2 \quad  , \quad
 \hat E_2 = i \sigma_2 \otimes \sigma_1 \quad  , \quad
 \hat E_3 = i \sigma_2 \otimes \sigma_3 \quad , &(2.4b)\cr}$$
 if we wish.

 Let $\gamma_\mu$ and $E_j$ be $2^{n+1} \times 2^{n+1}$ quaternionic IRS
 of some $C(p,q)$.  Since $E_1$, $E_2$, and $E_3$ generate the Clifford
 algebra $C(0,3)$ and since any representation of $C(0,3)$ is fully
 reducible, we can rewrite $E_j$'s as a diagonal matrix sum
 $$E_j = \pmatrix{\hat E_j & & &0\cr
  &\hat E_j & &\cr
  &  &\ddots&\cr
  0 & & &\hat E_j\cr} \ ,\ (j = 1,2,3)\eqno(2.5)$$
  if we choose a suitable basis.  Although $C(0,3)$ could have another
  inequivalent IRS with $\hat E_j$ being replaced by $-\hat E_j$, the
  latter cannot appear in the right side of Eq. (2.5) because $E_j$ must
  satisfy Eq. (1.8b).  We can then rewrite $\gamma_\mu$'s as
   $2^{n-1} \times 2^{n-1}$ block matrices
  $$\gamma_\mu = \pmatrix{M_{11}\ , \ &M_{12}\ ,\ &\dots\ ,\ &M_{1m}\cr
  \noalign{\vskip 5pt}%
  M_{21}\ ,\ &M_{22}\ ,\ &\dots\ ,\ &M_{2m}\cr
   \noalign{\vskip 5pt}%
  \dots&\dots&\dots&\dots\cr
   \noalign{\vskip 5pt}%
  M_{m1}\ ,\ &M_{m2}\ ,\ &\dots\ ,\ &M_{mm}\cr}\eqno(2.6)$$
 in terms of $4 \times 4$ matrices $M_{\alpha \beta}\ (\alpha , \beta =
 1,2, \dots , m)$ with $m = 2^{n-1}$.  Here, we have suppressed the extra
 index $\mu$ for simplicity.  The condition $[\gamma_\mu , E_j ] = 0$
 leads to $[\hat E_j , M_{\alpha \beta} ] = 0$ among $4 \times 4$
 matrices $\hat E_j$ and $M_{\alpha \beta}$.  We then apply the result of Eqs.
 (1.8) and (1.9) for the Clifford algebra $C(0,3)$ with identifications
 of $\gamma_\mu = \hat E_j$ and $E_j = \rho (e_j)$ to conclude that
 $M_{\alpha \beta}$ must have forms of
 $$M_{\alpha \beta} = a_{0 , \alpha \beta} \rho (e_0) + \sum^3_{j=1}
 a_{j, \alpha \beta} \rho (e_j) \eqno(2.7)$$
 for some real constants $a_{\mu , \alpha \beta}\ (\mu = 0, 1,2,3)$.
 Setting
 $$q_{\alpha \beta} = a_{0 , \alpha \beta} e_0 + \sum^3_{j=1}
 a_{j, \alpha \beta} e_j \quad .\eqno(2.8)$$
 Then Eq. (2.7) gives
 $$M_{\alpha \beta} = \rho (q_{\alpha \beta})$$
 in terms of real quaternion $q_{\alpha \beta}$.  Therefore, we may
  symbolically write
 Eq. (2.6) as
 $$\gamma_\mu = \rho (\Lambda_\mu ) \eqno(2.9)$$
 in terms of $2^{n-1} \times 2^{n-1}$ matrix $\Lambda_\mu$ over the real
 quaternion algebra
 $$\Lambda_\mu = \pmatrix{q_{11} &q_{12} &\dots &q_{1m}\cr
 \noalign{\vskip 5pt}%
 q_{21} &q_{22} &\dots &q_{2m}\cr
  \noalign{\vskip 5pt}%
  \dots &\dots &\dots &\dots\cr
   \noalign{\vskip 5pt}%
 q_{m1} &q_{m2} &\dots &q_{mm} \cr} \quad . \eqno(2.10)$$
 This establishes the desired relationship stated in the beginning of
 this section.  Also, Eq. (2.10) offers a particular example of the
 Wedderburn theorem [10] on a realization of irreducible modules in terms
 of a matrix algebra over some division algebras.

 Although we can recast  also almost complex representations similarly in
 terms of complex matrices
  with correspondence $i \leftrightarrow  \hat J =
  \pmatrix{0 &-1\cr
  1 &0\cr}$, we will not go into the details.

 \medskip

 \noindent {\bf  \underbar{Remark 2.1}}

 \medskip

 Somehow, quaternionic representation does not appear in physical
 applications.  However, a possibility for it exists.  Consider the Dirac
 equation
 $$\left( \gamma^\mu \partial_\mu - m \right) \psi = 0 \eqno(2.11)$$
 where we have set
 $$\gamma^\mu = \eta^{\mu \nu} \gamma_\nu \eqno(2.12)$$
 as usual in terms of the inverse flat metric $\eta^{\mu \nu}$ of
 $\eta_{\mu \nu}$.  Suppose that we are considering the case of the
 Majorana field so that $\gamma_\mu$ (and hence $\gamma^\mu$) are real
 matrices.  We can then impose the Majorana condition
 $$\psi^* = \psi \quad . \eqno(2.13)$$
 Since the Dirac equation should be compatible with Einstein's relation
 $p_0^2 = \underline{p}^2 + m^2$, the corresponding Clifford algebra must
 be $C(3,1)$ in general, which admits the well-known 4-dimensional normal
 IRS.  However for the special case of $m=0$, the Minkowski metric may be
 so chosen that the underlying Clifford algebra can be identified as
 $C(1,3)$ instead.  Note that $C(1,3)$ leads to the 8-dimensional real
 quaternionic representation.  Therefore, there exists in principle a
 possibility that we may have a 8-dimensional zero-mass Majorana field.
 One interesting feature of the quaternionic case is that the theory
 will automatically possess then an internal SU(2) symmetry generated
 by $E_j$'s.  In this connection, we recall the argument that Caianiello
 [3] used against the Majorana theory of neutrinos.  Suppose that the
 physics should be valid also in the 5-dimensional Kaluza-Klein
 space-time.  The resulting Clifford algebra would then become $C(4,1)$
 or $C(1,4)$ which admit only 8-dimensional
 real representation, corresponding
 to almost complex or quaternionic realization.  Note that only
 unphysical $C(3,2)$ allows the 4-dimensional realization for $N=5$.
 Therefore, if we require neutrinos to still  be 4-dimensional in the
 Kaluza-Klein theory, it cannot be Majorana.  Although this is not
 exactly the way Caianiello presented his argument in the paper [3], it
 is essentially related to the discussion we utilized here.

 \medskip

 \noindent {\bf \underbar{Remark 2.2}}

 \medskip

 Any quaternionic IRS can have a natural gauge theory in a sense that the
 Dirac equation (2.11) can be gauged into
 $$\left\{ \gamma^\mu \left( \partial_\mu - e \sum^3_{a=1} E_a A^a_\mu
 \right) + m \right\} \psi =0 \eqno(2.14)$$
 by introducing the Yang-Mills SU(2) gauge field $A^a_\mu \ (a = 1,2,3)$.

 \medskip

 \noindent {\bf 3. \underbar{Orthogonality Relation and Fierz
 Transformations}}

 \medskip

 Let $\gamma_\mu$ be the $d \times d$ real matrices as before.  We
 construct $2^N$ real matrices $\Gamma_A\ (A = 1,2, \dots , 2^N)$ by
 $$\Gamma_A = E, \gamma_\mu , \gamma_\mu  \gamma_\nu (\mu < \nu),
 \gamma_\mu \gamma_\nu \gamma_\lambda (\mu < \nu < \lambda), \dots ,
 \gamma_1 \gamma_2 \dots \gamma_N \quad . \eqno(3.1)$$
 First, we note
 $$(\Gamma_A )^2 = \pm E \eqno(3.2)$$
 for any $\Gamma_A$, so that $\Gamma_A^{-1} = \pm \Gamma_A$.  Second, we
 note the validity of
 $${\rm Tr}\ \Gamma_A = 0 \eqno(3.3)$$
 for all $\Gamma_A$'s except for the following special cases:
 \item{(i)} $N= 2n =$ even,
 $$\Gamma_A = E \eqno(3.4a)$$
 \item{(ii)} $N= 2n +1 =$ odd,
 $$\Gamma_A = E  \quad , \quad {\rm and}\quad \Gamma_A = \gamma_1
 \gamma_2 \dots \gamma_N \quad . \eqno(3.4b)$$

 \noindent This is due to the
 following reason; except for cases specified by Eqs.
 (3.4), we can always find some $\Gamma_B$ satisfying
 $$\Gamma_B \Gamma_A = - \Gamma_A \Gamma_B $$
 so that we calculate
 $${\rm Tr}\ \Gamma_A = - {\rm Tr}\ \left( \Gamma^{-1}_B \Gamma_A
 \Gamma_B \right) = - {\rm Tr}\ \Gamma_A = 0 \quad .$$
 For example, if $\Gamma_A = \gamma_1$, then we can choose $\Gamma_B =
 \gamma_2$, assuming $N  \geq 2$.

 Next, for any two $\Gamma_A$ and $\Gamma_B$, we can always find the 3rd
 $\Gamma_C$ satisfying
 $$\Gamma_A \Gamma_B = \epsilon_{AB} \Gamma_C \quad , \quad \epsilon_{AB}
 = \pm 1 \quad . \eqno(3.5)$$
 Moreover, for a fixed $\Gamma_B$, a set consisting of all $\Gamma_A
 \Gamma_B$ covers the original set given by Eq. (3.1) except possibly for
 individual signs, when we change $\Gamma_A$.  For the case of $N = 2n =$
 even, these facts are sufficient to prove that $2^N$ matrices $\Gamma_A$
 are linearly independent just as in the complex Clifford case [5].

 Now, the orthogonality relations [11] are given by

 \medskip

 \noindent {\bf (I) \underbar{Normal Representation}}

 \medskip

 $$\sum_A \left( \Gamma^{-1}_A \right)_{jk} (\Gamma_A)_{\ell m} = {2^N
 \over d}\ \delta_{jm} \delta_{\ell k} \eqno(3.6)$$

 \medskip

 \noindent {\bf (II) \underbar{Almost Complex Representation}}

 \medskip

 $$\sum_A \left( \Gamma^{-1}_A \right)_{jk} (\Gamma_A )_{\ell m} =
 {2^N \over d}\ \left\{ \delta_{jm} \delta_{\ell k} - J_{jm} J_{\ell k}
 \right\}\eqno(3.7)$$

 \vfill\eject

 \medskip

 \noindent {\bf  (III) \underbar{Quaternionic Representation}}

 \medskip

 $$\sum_A \left( \Gamma^{-1}_A \right)_{jk} (\Gamma_A )_{\ell m} =
 {2^N \over d}\ \left\{ \delta_{jm} \delta_{\ell k} - \sum^3_{a=1}
 (E_a)_{jm} (E_a)_{\ell k} \right\} \eqno(3.8)$$
 for all $j,k,\ell,m = 1,2,\dots ,d$.
  To illustrate, let us prove Eq. (3.7) for the almost complex case.

 Let $Y$ be an arbitrary $d \times d$ real matrix and set
 $$S = \sum_A \Gamma^{-1}_A Y \Gamma_A \quad .
 \eqno(3.9)$$
 We calculate then
 $$\Gamma_B^{-1} S \Gamma_B = \sum_A \left( \Gamma_A \Gamma_B
 \right)^{-1} Y \left( \Gamma_A \Gamma_B \right) $$
 and change the summation variable from $\Gamma_A$ to $\Gamma_C$ as in
 Eq. (3.5) to find
 $$\Gamma^{-1}_B S \Gamma_B = \sum_C \Gamma^{-1}_C Y
 \Gamma_C = S \quad .$$
 In other words, we find
 $$[ S, \Gamma_B ] = 0 \eqno(3.10)$$
 for any $\Gamma_B$. For the case of the almost complex representation,
 $S$ must assume the form of  Eq. (1.6), i.e.,
 $$S = \sum_A \Gamma^{-1}_A Y \Gamma_A = a E + b J \eqno(3.11)$$
 for some real constants $a$ and $b$.  We note that we have
 $${\rm Tr}\ J = 0\eqno(3.12)$$
 because of the following reason.  First, Tr $J$ must clearly be real,
 since $J$ is a real matrix.  Second, extending the field from the real
 to complex field, we regard $J$ to be a complex matrix which happens to
 be real.  But then eigenvalues of $J$ must be $\pm i$ in view of Eq.
 (1.7a), i.e. $J^2 = -E$.  Therefore, Tr $J$ is purely imaginary, unless
 it is identically zero.  These two facts prove Eq. (3.12). We can now
 determine $a$ and $b$ in Eq. (3.11) as follows.  Taking the trace of
 both sides of Eq. (3.11), it gives
 $$a = {2^N \over d}\ {\rm Tr}\ Y \quad .$$
 Next, we multiply $J$ to both sides of Eq. (3.11) and take the trace.
 When we note $[J , \Gamma_A ] = 0$ in view of Eq. (1.7b), we will obtain
 $$b = - {2^N \over d}\ {\rm Tr}\ (JY) \quad .$$
 Inserting these results to the right side of Eq. (3.11) and noting that
 the $d \times d$ matrix $Y$ is arbitrary, this leads to the desired
 orthogonality relation Eq. (3.7).

 When we set $m= \ell$ and sum over the values of $1,2, \dots , d$ and
 note Eq. (3.3), then these orthogonality relations Eqs. (3.6)-(3.8) will
 lead to the dimensionality part of the theorem stated in Eqs.
 (1.11), (1.15),  and (1.19).  See  ref. [11] for details.

 The orthogonality relation Eq. (3.6)
 for the normal representation
  has the same form as that for the
 complex case [13].  Especially, any $d \times d$ matrix $Y$ can be
 expanded as
 $$\eqalignno{Y &= \sum_A a_A \Gamma_A \quad , &(3.13a)\cr
 \noalign{\vskip 4pt}%
 a_A &= {d \over 2^N}\ {\rm Tr}\ \big( \Gamma^{-1}_A Y \big)
 \quad , &(3.13b)\cr}$$
 by multiplying $Y_{kj}$ to both sides of Eq. (3.6).  However, such a
 expansion is \underbar{not} possible for both almost complex and
 quaternionic IRS.  In other words,
 the Burnside theorem will not hold for these cases.  Similarly, the
 standard Fierz transformation ([13], [14]) will work only for the normal
 representation.  In order to see them more clearly, we will consider the
 following modified orthogonality relations:

 \medskip

 \noindent {\bf (II$^\prime$) \underbar{Almost Complex Representation}}

 \medskip

 $$\sum_A \left\{ \left( \Gamma^{-1}_A \right)_{jk}
  \left( \Gamma_A \right)_{\ell m} +
 \left( \Gamma^{-1}_A D^{-1} \right)_{jk}
 \left( D \Gamma_A\right)_{\ell m}\right\} =
 {2^{N+1} \over d}\ \delta_{jm} \delta_{\ell k}\eqno(3.14)$$

 \medskip

 \noindent {\bf (III$^\prime$) \underbar{Quaternionic Representation}}

 \medskip

   $$\sum_A \left\{ \left( \Gamma^{-1}_A \right)_{jk}
  \left( \Gamma_A \right)_{\ell m} -
 \sum^3_{a=1}
 \left( \Gamma^{-1}_A E_a  \right)_{jk}
 \left( E_a  \Gamma_A\right)_{\ell m} \right\}=
 {2^{N+2} \over d}\ \delta_{jm} \delta_{\ell k} \quad .\eqno(3.15)$$
 For the almost complex representation, we set
 $$S = \sum_A \left\{ \Gamma^{-1}_A Y \Gamma_A + \Gamma^{-1}_A D^{-1} Y D
 \Gamma_A
 \right\}$$
 for an arbitrary real $d \times d$ matrix $Y$,
  instead of Eq. (3.9) and
 repeat the same procedure by noting $JD + DJ = 0$ in view of Eq.
 (1.17a) to obtain Eq. (3.14).
  We remark that Eq. (3.14) also results from the orthogonality
 relation Eq. (3.6) for the normal representation of $C(p+1,q)$ or
 $C(p,q+1)$  by Remark 1.3.  Especially, Eq. (3.13) must be replaced
 now by
 $$\eqalignno{Y &= \sum_A \big\{ a_A \Gamma_A + b_A D \Gamma_A \big\}
 &(3.16a)\cr
 a_A &= {d \over 2^{N+1}} \ {\rm Tr}\ \big( \Gamma^{-1}_A Y \big) \quad ,
 \quad b_A = {d \over 2^{N+1}} \ {\rm Tr}\ \big( \Gamma^{-1}_A D^{-1} Y
 \big) \quad . &(3.16b)\cr}$$
 Similarly for the quaternionic case, the expansion of any $d \times d$
 matrix $Y$ will be expressed in the form of
 $$Y = \sum_A \left\{ a_{A,0} \Gamma_A + \sum^3_{j=1} a_{A,j} E_j
 \Gamma_A \right\} \eqno(3.17)$$
 for some constants $a_{A,0}$ and $a_{A,j}\  (j=1,2,3)$.

 To prove Eq. (3.15), we set
 $$S = \sum_A \left\{ \Gamma^{-1}_A Y \Gamma_A - \sum^3_{a=1}
 \Gamma^{-1}_A E_a Y E_a \Gamma_A \right\} \quad .\eqno(3.18)$$
 We can then prove again the validity of Eq. (3.10) so that
 $$S = a_0 E + \sum^3_{j=1} a_j E_j \eqno(3.19)$$
 for some constants $a_0$ and $a_j$.  Taking the trace of both sides, and
 noting Tr $E_j = 0$ by the same reasoning as in the proof of Eq. (3.12),
 it leads to
 $$a_0 = {2^{N+2} \over d}\ {\rm Tr}\ Y \quad .$$
 Next, we calculate $a_j \ (j=1,2,3)$ by multiplying
  $E_j$ to Eqs. (3.19) and
 (3.18), and note $[E_j,\Gamma_A ] =0$ in view of Eq. (1.8a).  Utilizing
 Eq. (1.8b), this leads to
 $$a_j = -{2^N \over d}\ \left\{ {\rm Tr}\ (E_j Y) - \sum^3_{a=1} \ {\rm
 Tr}\ (E_a E_j E_a Y) \right\} = 0$$
 since we have
 $$\sum^3_{a=1} E_a E_j E_a = E_j \quad .$$
 Therefore, we find
 $$S = {2^{N+2}\over d}\ ({\rm Tr}\ Y) E \quad .$$
 Comparing this with Eq. (3.18) and noting that the $d \times d$ matrix
 $Y$ is arbitrary, we obtain the desired result of Eq. (3.15).

 We are now in position to discuss the Fierz identities.  Let $\psi_1,\
 \psi_2,\ \psi_3,$ and $\psi_4$ be the mutually anticommuting Grassmann
 $d$-component spinors on which $\gamma_\mu$  act.  Multiplying
 $(\psi_1 C)_j (\psi_2)_k (\psi_3 C)_\ell (\psi_4)_m$
 to both sides of Eq. (3.6) and summing over $j$, $k$, $\ell$, and $m$,
 we obtain the Fierz identity ([13], [14]) for the normal representation:
 $$\sum_A \left( \psi_1 C \Gamma^{-1}_A \psi_2 \right) \cdot \left(
 \psi_3 C \Gamma_A \psi_4 \right) = - {2^N \over d}\ \left( \psi_1 C
 \psi_4\right) \cdot \left( \psi_3 C \psi_2 \right)
 \quad , \eqno(3.20)$$
 where we have written for simplicity
 $$( \psi Q \phi ) = \sum^d_{j,k=1} \psi_j Q_{jk} \phi_k \quad .
 \eqno(3.21)$$
 Since $\psi_2$ and $\psi_4$ are arbitrary spinors, we may replace them
 by $\Gamma^{-1}_B \psi_2$ and $\Gamma_B \psi_4$, respectively to find
  $$\sum_A \left( \psi_1 C \Gamma^{-1}_A  \Gamma^{-1}_B
  \psi_2 \right) \cdot \left(
 \psi_3 C \Gamma_A \Gamma_B  \psi_4 \right) =
 - {2^N \over d}\ \left( \psi_1 C \Gamma_B
 \psi_4\right) \cdot \left( \psi_3 C
 \Gamma_B^{-1} \psi_2 \right) \quad . \eqno(3.22)$$
 When we sum over restricted ranges over $B$ (for example $\Gamma_B
 = \gamma_\mu$), this leads to the standard Fierz identity
 for the normal or
 complex IRS (see [13] and [14]).
 When we note (see Eq. (2.12) for the definition of $\gamma^\mu$)
 $$(\gamma_\mu )^{-1} = \gamma^\mu \quad , \eqno(3.23)$$
 then $\Gamma^{-1}_A$ for $\Gamma_A$'s given by Eq. (3.1) is written as
 $$\left( \Gamma_A \right)^{-1} = E, \gamma^\mu , \gamma^\nu \gamma^\mu
 (\mu < \nu), \gamma^\lambda \gamma^\nu \gamma^\mu (\mu < \nu <
 \lambda), \dots , \gamma^N \gamma^{N-1} \dots \gamma^1 \quad .
 \eqno(3.24)$$
 Therefore, relations Eqs. (3.20) and (3.22) are invariant under
 SO($p,q$) transformation
 $$\psi_j \rightarrow U \psi_j \quad (j = 1,2,3,4)\eqno(3.25)$$
 for $U$ given by Eq. (1.30).

 These relations are also valid for any complex Clifford algebras.  In
 this connection, we remark that Caianiello in papers [1] and [2] had
 utilized the special case of these Fierz transformations for $C(3,1)$
 for his study of 4-Fermi weak interactions whose explicit V-A form was
 still unknown at that time.

 For quaternionic representations, the situation is, however, quite
 different.  From Eqs. (3.8) and (3.15), we find
 $$\eqalignno{\sum_A \big( &\psi_1 C \Gamma^{-1}_A \psi_2 \big) \cdot
 \big( \psi_3 C \Gamma_A \psi_4 \big)\cr
 &=- {2^N \over d}\ \bigg\{ \big( \psi_1 C \psi_4 \big) \cdot \big(
 \psi_3 C \psi_2 \big) - \sum^3_{a=1} \big( \psi_1 C E_a \psi_4 \big)
 \cdot \big( \psi_3 C E_a \psi_2\big) \bigg\}\quad ,&(3.26)\cr
 \sum_A \bigg\{ \big( &\psi_1 C \Gamma^{-1}_A \psi_2 \big) \cdot
 \big( \psi_3 C \Gamma_A \psi_4 \big) -
 \sum^3_{a=1} \big( \psi_1 C \Gamma_A^{-1} E_a  \psi_2\big)
 \cdot \big( \psi_3 C \Gamma_A E_a \psi_4\big) \bigg\}\cr
&=- {2^{N+2} \over d}\ \big( \psi_1 C \psi_4 \big) \cdot \big(
 \psi_3 C \psi_2 \big) \quad . &(3.27)\cr}$$

\medskip

\noindent {\bf \underbar{Remark 3.1}}

\medskip

For complex realization of the Clifford algebras, it is more customary to
use symbols $\overline \psi_1$ and $\overline \psi_3$ instead of
$\psi_1 C$ and $\psi_3 C$ as in here.

\medskip

\noindent {\bf 4. \underbar{Octonionic Triple System and Octonion}}

\medskip

As an application of the Fierz transformations, we will construct in this
section the real as well as complex octonion algebras.  For this, we
need to introduce the notion of the octonionic triple system [15].  Let
$V$ be a $N$-dimensional vector space over the real or complex field

$$N = \ {\rm Dim}\ V \quad . \eqno(4.1)$$
We suppose that $V$ possesses a symmetric bilinear non-degenerate form
$<x|y>$ for $x,y\ \epsilon\ V$ so that we have especially $<y|x>\  =
\ <x|y>$.  Moreover, we assume the existence of the triple product
$[x,y,z]$, $$[x,y,z] \ :\ V \otimes V \otimes V \rightarrow V \eqno(4.2)$$
satisfying the following conditions:
\item{(i)} $[x,y,z]$ is totally antisymmetric in $x,y,z\ \epsilon\ V$
\hfill (4.3a)
\smallskip
\item{(ii)} $<w|[x,y,z]>$ is totally antisymmetric in 4 variables
$x,y,z, w\ \epsilon\ V$ \hfill (4.3b)
\smallskip
\item{(iii)} $<[x,y,z]|[u,v,w]>$
\vskip -.5cm
$$ = \alpha \sum_P (-1)^P <x|u><y|v><z|w>
\qquad\qquad\qquad\qquad\qquad\qquad\qquad\qquad \eqno(4.3c)$$
\vskip -.5cm
$$\ + \ {\beta \over 4}\ \sum_{P,P^\prime} (-1)^P
(-1)^{P^\prime} <x|u><y|[z,v,w]>\quad ,
\qquad\qquad\qquad\qquad\qquad\qquad\qquad$$

\noindent for some  constants $\alpha$ and $\beta$, where the summations in Eq.
(4.3c) are over 3! permutations $P$ and $P^\prime$ of $x,\ y$, and $z$
and of $u,$ $v,$ and $w,$ respectively.

In view of the non-degeneracy of $<x|y>$,  Eq. (4.3c) is actually
equivalent to the triple product equation
$$\eqalign{[[&x,y,z],u,v]\cr
\noalign{\vskip 4pt}%
&= {1 \over 2} \sum_P (-1)^P \big\{ \alpha [<y|v><z|u> -<y
|u><z|v>] - \beta <u|[v,y,z]>\big\}x\cr
\noalign{\vskip 4pt}%
&\quad - {1 \over 2} \ \beta \sum_P (-1)^P \big\{ <x|v>[u,y,z]\  + <x|u>
[v,z,y]\big\} \quad . \cr}\eqno(4.4)$$
In [15], we have proved that the solutions of Eqs. (4.3) are possible
only for two cases of
\item{(a)} $N=8$ with $\alpha = \beta^2 \not= 0$ \hfill (4.5a)

\item{(b)} $N=4$ with $\beta = 0\ , \ \alpha \not= 0$ \hfill (4.5b)

\noindent if we ignore the uninteresting case of $\alpha = \beta =0$.  We
named two cases as octonionic and quaternionic triple systems,
respectively in [15].  Now we will concentrate on the first case of
$N=8$.  If we change the normalizations of $[x,y,z]$ and/or $<x|y>$, then
we can normalize $\alpha$ and $\beta$ to be given by
$$\alpha = 1 \quad , \quad \beta = -1 \eqno(4.6)$$
which we will assume hereafter.  Since $<x|y>$ is nondegenerate, we can
find a element $e\ \epsilon\  V$ satisfying
$$<e|e> \ = 1 \quad . \eqno(4.7)$$
For any such $e\ \epsilon\ V$, we define a bilinear product $xy$ in $V$ by
$$xy = [x,y,e]\  + <x|e> y \ +<y|e>x\  - <x|y>e \quad . \eqno(4.8)$$
We have clearly then
$$xe = ex = x \eqno(4.9)$$
\noindent from Eq. (4.3a) for any $x\ \epsilon\ V$ so that $e$ is the unit
element.  Moreover, from Eqs. (4.3b) and (4.3c), we can easily verify the
composition law
$$<xy|xy> \ =\  <x|x><y|y> \quad . \eqno(4.10)$$
Therefore, the resulting algebra must be an octonion algebra by the
Hurwitz theorem [16].  If the underlying field is real (or complex), then
it gives a real (or complex) octonion algebra.  Conversely, the
octonionic triple product can be expressed in terms of the octonion
algebra by
$$\eqalign{[x,y,z] = \ &{1 \over 2}\ \{ (x,y,z)\  + <x|e>[y,z]\  + <y
|e>[z,x]\cr
&\quad + <z|e>[x,y] \ -<z|[x,y]>e\}\cr}\eqno(4.11a)$$
where $(x,y,z)$ and $[x,y]$ are associator and commutator, i.e.
$$\eqalign{&(x,y,z) = (xy)z - x(yz)\cr
&[x,y] = xy - yx \quad .\cr}\eqno(4.11b)$$
For details, see ref. [15].  Historically the relations Eqs. (4.11) have
  been essentially   discovered  by de Wit and Nicolai
[17] and by G$\ddot {\rm u}$rsey and Tze [18], although they did not use the
terminology of the triple product.  Also, in [15] we proved that the
SO(7) spinor space will lead to the octonionic triple system, although we
did not make  its explicit construction.

Our remaining task is a concrete realization of the octonionic system out
of the 8-~dimensional IRS of the Clifford algebra $C(p,q)$ with $p+q =7$.
For the complex case, we can utilize any $C(p,q)$.  However, only
$C(0,7)$ and $C(4,3)$ can lead to the 8-dimensional normal
representations for the real case, so that we have to restrict ourselves
only to these two cases for the construction of real octonions.

In what follows, we will consider only the real cases of $C(0,7)$ and
$C(4,3)$ unless it is stated otherwise.  Let $V$ be the real IRS, whose
elements are now 8-dimensional real spinors on which
$\gamma_\mu$'s act.  From the results of section 1, there exists a real $8
\times 8$
charge conjugation matrix $C$ satisfying
$$\eqalignno{&C \gamma_\mu C^{-1} = -\gamma^T_\mu \quad (\mu =1,2,\dots
,7)&(4.12a)\cr
&C^T = C \quad . &(4.12b)\cr}$$
We then introduce the bilinear product $<x|y>$ in $V$ by
$$<x|y> \ = (x C y) \eqno(4.13)$$
in the notation of the previous section as in Eq. (3.21).  However, all
spinors in $V$ are here assumed to be $C$-numbers i.e. they commute
(rather than anticommute) with each other.  Then, $<x|y>$ is a real
non-degenerate symmetric bilinear form in $V$.

We now construct the triple product in $V$ by
$$[x,y,z] = {1\over 3}\ \sum^7_{\mu=1} \big\{ \gamma_\mu x (y C
\gamma^\mu z) + \gamma_\mu y(zC\gamma^\mu x) +\gamma_\mu z(xC \gamma^\mu
y)\big\}\quad .\eqno(4.14)$$
\noindent From Eqs. (4.12), it is simple to verify
$$(x C \gamma^\mu y ) = -(y C \gamma^\mu x) \eqno(4.15)$$
to be antisymmetric in $x$ and $y$, since $x$ and $y$ are mutually
commuting $C$-number spinors.  Then, $[x,y,z]$ defined by Eq. (4.14) is
totally antisymmetric in $x,$ $y,$ and $z$, satisfying the condition Eq.
(4.3a).  We next calculate
$$\eqalign{<w|&[x,y,z]>\cr
&= {1 \over 3} \sum^7_{\mu=1} \big\{ (w C \gamma_\mu x )(yC \gamma^\mu z)
+ (w C \gamma_\mu y)(zC\gamma^\mu x)
+ (w C \gamma_\mu z) (x C \gamma^\mu y)\big\} \cr}\eqno(4.16)$$
which is again totally antisymmetric in 4-variables $x,$ $y,$ $z,$ and
$w$ because of Eq. (4.15).  We have yet to verify the validity of Eq.
(4.3c), whose proof requires the Fierz identity of section 3.  However,
since we are discussing $C$-number spinors, the sign on the right sides
 of Eqs. (3.20) and (3.22) must be reversed now.  After some calculations
 (see Appendix) we can then rewrite Eq. (4.14)
 (with Einstein's summation convention for repeated indices) also as
 $$\eqalign{[x,y,z] &= \gamma_\mu x (y C \gamma^\mu z) + y<z|x> -\  z<x|y>\cr
  &= \gamma_\mu y (z C \gamma^\mu x) + z<x|y> - \ x<y|z>\cr
  &= \gamma_\mu z (x C \gamma^\mu y) + x<y|z> - \ y<z|x>
  \quad . \cr}\eqno(4.17)$$
  Here, we used the fact that we have
  $$(C Q C^{-1})^T = CQ C^{-1}\eqno(4.18a)$$
  for $Q=E$ and $Q = \gamma_\mu \gamma_\nu \gamma_\lambda\
  (\mu < \nu < \lambda)$ but we have
  $$(C Q C^{-1} )^T = - C Q C^{-T}\eqno(4.18b)$$
  for $Q=\gamma_\mu$, and $Q= \gamma_\mu \gamma_\nu\ (\mu < \nu)$.  Since
  we have $\gamma_1 \gamma_2 \gamma_3 \gamma_4 \gamma_5
  \gamma_6 \gamma_7 = \pm E$
  by Eq. (1.13), the summation on $\Gamma_A$ in the Fierz
  identity can be rewritten in terms of only $E,\gamma_\mu, \gamma_\mu
  \gamma_\nu \ (\mu < \nu)$ and $\gamma_\mu \gamma_\nu \gamma_\lambda \
  (\mu < \nu < \lambda)$.  We remark here that the explicit form of the
  Fierz identity for the normal representations can be more easily
  computed from the formula given by Braden [14] rather than the direct
  use of Eqs. (3.20) and (3.22).

  On the basis of Eq. (4.17), we calculate first
  $$\eqalign{<[&u,v,w]|[x,y,z]>\cr
   &= \ <[u,v,w]|x> <y|z> -< [u,v,w]|y><x|z>\cr
   &\quad -<z|w>(uC\gamma_\mu v)(xC \gamma^\mu y)\  - <v|w> (z C \gamma_\mu
   u)(xC \gamma^\mu y)\cr
   &\quad + <u|w>(zC \gamma_\mu v)(xC \gamma^\mu y)\cr
   &\quad - {1 \over 2}\  (z C [\gamma_\mu ,
   \gamma_\nu ]w)(uC \gamma^\nu v) (xC \gamma^\mu y)\quad .\cr}$$
   Interchanging $z$ and $w$,  adding both,
    and noting Eq. (4.17), then the expression
   $$K(u,v,w|x,y,z) = \ <[u,v,w]|[x,y,z]> + <[u,v,z]|[x,y,w]> \eqno(4.19)$$
   is rewritten as

   \smallskip

   $$\eqalign{K(u,&v,w|x,y,z) = -2 <z|w>
   \{<u|[x,y,v]> - <u|x><y|v> + <u|y><x|v>\}\cr
   &+ <y|z><x|[u,v,w]> - <x|z><y|[u,v,w]>\cr
    &+ <y|w><x|[u,v,z]> - <x|w><y|[u,v,z]>\cr
    &+ <v|w>\{ <u|[x,y,z]> + <z|x><y|u> -
    <z|y> <x|u>\}\cr
     &- <u|w>\{ <v|[ x,y,z]> + <z|x><y
     |v> - <z|y><x|v>\}\cr
       &- <v|z>\{ <w|[ x,y,u]> + <w|y><x
     |u> - <w|x><y|u>\}\cr
         &+ <u|z>\{ <w|[ x,y,v]> + <w|y><x
     |v> - <w|x><y|v>\} \quad . \cr}\eqno(4.20)$$
     On the other side, the second term in Eq. (4.19) can be rewritten as

     \smallskip

     $$\eqalign{<[&u,v,z]|[x,y,w]>\  = <[z,u,v]|[w,x,y]>\cr
     &=- <[z,u,y]|[w,x,v]> +\  K(z,u,v|w,x,y)\cr
      &=- <[z,y,u]|[w,v,x]> +\  K(z,u,v|w,x,y)\cr
     &=- \{ -<[z,y,x]|[w,v,u]> +\  K(z,y,u|w,v,x)\} +K(z,u,v|w,x,y)\cr
     &= <[x,y,z]|[u,v,w]> -\  K(z,y,u|w,v,x) + K(z,u,v|w,x,y) \quad ,\cr}$$

     \smallskip

     \noindent so that Eq. (4.19) leads to

     \smallskip

     $$\eqalign{2<[&u,v,w]|[x,y,z]>\cr
     &= K(u,v,w|x,y,z) + K(z,y,u|w,v,x) - K(z,u,v|w,x,y)\quad .
     \cr}\eqno(4.21)$$

     \smallskip

     \noindent We can
     verify that Eqs. (4.21) and (4.20) give the desired relation
     Eq. (4.3c) with $\alpha = - \beta =1$.  This completes the proof
     that $[x,y,z]$ defines the desired octonionic triple system. For
     $C(0,7)$, we can choose $C=E$ by the remark 1.4.  Then, we find
     $<x|x> \not= 0$ if $x \not= 0$.  This is sufficient to prove that
     the resulting octonion algebra is a real division algebra [19].

     On the other side, we can choose $C = \gamma_5 \gamma_6 \gamma_7$
     for $C(4,3)$ so that we have $C^2 =E$ and Tr $C=0$.  Especially, $C$
     can have equal numbers of eigenvalues 1 and $-1$.  Then, $<x|x>$ can
     be zero even for non-zero $x$, implying that the resulting octonion
     is the split Cayley algebra.

     We remark that the octonionic triple system has been used
      also to obtain
     a solution of the Yang-Baxter equation in ref. [15] and in [20].

     \medskip

     \noindent {\bf 5. \underbar{Comment on Composition Algebra}}

     \medskip

     As another application of the Clifford algebra, we will discuss
     dimensional part of the Hurwitz theorem [16] in this section.
     Consider the composition law Eq. (4.10), i.e.
     $$<xy|xy>\  =\  <x|x><y|y> \eqno(5.1)$$
     for a bilinear symmetric non-degenerate form $<x|y>$.  It is well
     known that the dimension of the vector space $V$ is restricted to 1,
     2, 4, or 8, provided that $V$ is finite dimensional.  However, $V$
     could be infinite dimensional [21], if it has no unit element.
     Assuming the existence of the unit element $e$, Eq. (5.1) is known
     ([16], [19]) to lead to the validity of
     $$\eqalignno{&(xy) \overline y =\  <y|y>x \quad , &(5.2)\cr
     &<xy|z>\  = \ <x | z \overline y> &(5.3)\cr}$$
     where $\overline x$ is the conjugate of $x$, defined by
     $$\overline x = 2 < x|e> e-x \quad . \eqno(5.4)$$
     Conversely, if Eqs. (5.2) and (5.3) hold valid, we will have the
     composition law Eq. (5.1) since we calculate
     $$<xy|xy> \ = \ <x|(xy)\overline y>\  = \
     <x|<y|y>x>\  = \ <y|y><x|x> \quad .$$

     We will show first that the dimension of any finite dimensional
     algebra satisfying Eq. (5.2) but not necessarily Eq. (5.3) must be
     limited to 1, 2, 4, or 8.  To be definite, we will consider only the
     case of $V$ being the real vector space.  Let $N=$ Dim $V$ be its
     dimension.  Then, the standard reasoning based upon non-degeneracy
     of $<x|y>$ implies [19] the existence of basis vectors $e_\mu\ (\mu
     = 0,1,2, \dots , N-1)$ with $e_0 =e$ satisfying
     $$<e_\mu | e_\nu >\  = \eta_{\mu \nu} = \cases{0\ , &if $\mu \not=
     \nu$\cr
     \noalign{\vskip 5pt}%
     1\ , &if $\mu = \nu =0,1,2, \dots , p-1$\cr
     \noalign{\vskip 5pt}%
     -1\ , &if $\mu = \nu =p, p+1, \dots ,N$\cr} \quad . \eqno(5.5)$$
     Since $<e|e>\  =1$, we must have $p \geq 1$.

     Introducing the right multiplication operator $R_y\ :\ V \rightarrow
     V$ by
     $$R_y x = xy \quad , \eqno(5.6)$$
     Eq. (5.2) is rewritten as
     $$R_{\overline  y} R_y = <y|y> I \eqno(5.7)$$
     where $I$ is the identity map in $V$.  Linearlizing Eq. (5.7) by
     letting $y \rightarrow y \pm z$, this leads to
     $$R_{\overline y} R_z + R_{\overline z} R_y = 2<y|z> I \quad .
     \eqno(5.8)$$
     Now, setting
     $$V_0 = \{ x| <x|e> = 0 \ ,\ x\ \epsilon\ V \} \eqno(5.9)$$
     then $V_0$ is spanned by $N-1$ basis vectors $e_1,e_2, \dots , e_{N-1}$,
     and hence
     $${\rm Dim}\ V_0 = N-1 \quad . \eqno(5.10)$$
     Choosing $y = e_j$ and $z = e_k$ for $j,k = 1,2, \dots , N-1$,
     Eq. (5.8) is rewritten as
     $$R_j R_k + R_k R_j = -2 \eta_{jk} I \ (j,k =1,2\dots , N-1)
     \eqno(5.11)$$
     where we have for simplicity set
     $$R_j = R_{e_j} \quad . \eqno(5.12)$$
     Clearly, Eq. (5.11) with Eq. (5.5) defines the Clifford algebra
     $C(q,p-1)$ by setting
     $$q = N -p \quad . \eqno(5.13)$$
     Since $R_j \ :\ V \rightarrow V$ with Dim $V=N$ can be identified
     with $N \times N$ matrices, it may be regarded as a $N \times N$
     matrix realization of the Clifford algebra.  Therefore, we must have
     $$N = md \eqno(5.14)$$
     for the dimension $d$ of the IRS of $C(q,p-1)$ where $m$ is the
     multiplicity of the irreducible components contained in $V$.

     Let
     $$N- 1 = 2n \quad {\rm or}\quad 2n+1 \quad . \eqno(5.15)$$
     Then, the dimension $d$ of $C(q,p-1)$ is given by
     $$d = 2^n \eqno(5.16)$$
     for the normal representation and by
     $$d = 2^{n+1} \eqno(5.17)$$
     for both almost complex and quaternionic representation.  Consider
     first the case Eq. (5.16) of the normal representation.  Solutions
     of Eqs. (5.14), (5.15) and (5.16) are then possible only for the
     following 4 cases of
     $$\eqalign{&{\rm (i)}\quad N=1 \quad , \quad d =1 \quad , \quad m=1
     \qquad\qquad\qquad\qquad\qquad\cr
     &{\rm (ii)}\quad N=2 \quad , \quad d=1 \quad , \quad m=2
      \qquad\qquad\qquad\qquad\qquad\cr
     &{\rm (iii)}\quad  N=4 \quad , \quad d=2 \quad , \quad m=2
      \qquad\qquad\qquad\qquad\qquad\cr
     &{\rm (iv)} \quad N=8 \quad ,\quad d=8 \quad , \quad m=1 \ \ \ .
      \qquad\qquad\qquad\qquad\qquad \cr} \eqno(5.18)$$
      For both almost complex and quaternionic realization, we must have
      Eq. (5.17) so that the solutions are limited to
      $$\eqalign{&{\rm (i)}\quad N=2 \quad , \quad d=2 \quad , \quad m=1
       \qquad\qquad\qquad\qquad\qquad\cr
       &{\rm (ii)}\quad N=4 \quad , \quad d=4 \quad , \quad m = 1
       \ \ \ . \qquad\qquad\qquad\qquad\qquad \cr}\eqno(5.17)$$
        Especially, the case of $N=8$ requires $C(q,p-1)$ with $p+q = 8$
        to be of the normal representation which is possible only for two
        cases of $C(0,7)\ (p=8,\  q=0)$ and $C(4,3) \ (p=q=4)$.  Just as in
        the previous section, these correspond to the real division
        octonion algebra and real split Cayley algebra, respectively.

     \medskip

     \noindent {\bf \underbar{Acknowledgement}}

     \medskip

     This paper is supported in part by the U.S. Department of Energy
     Grant No. DE-~FG-02-91ER40685.

     \vfill\eject

     \noindent {\bf \underbar{Appendix: Proof of Eq. (4.17)}}

     \medskip

     The Fierz-transformation for normal representations of $N=7$
     Clifford algebra is given as follows.  Setting
     $$\eqalignno{S &= \big( \psi_1 C \psi_2 \big) \big( \psi_3 C \psi_4
     \big) &(A.1)\cr
     V &= \sum^7_{\mu =1} \big( \psi_1 C \gamma_\mu \psi_2 \big) \big(
     \psi_3 C \gamma^\mu \psi_4 \big) &(A.2)\cr
     T &= - \sum^7_{\mu < \nu} \big( \psi_1 C \gamma_\mu \gamma_\nu
     \psi_2 \big)\big( \psi_3  \gamma^\mu \gamma^\nu \psi_4 \big) &(A.3)\cr
     W &= -\sum^7_{\mu < \nu < \lambda} \big( \psi_1 C \gamma_\mu
     \gamma_\nu
     \gamma_\lambda \psi_2 \big) \big( \psi_3 C \gamma^\mu \gamma^\nu
     \gamma^\lambda \psi_4 \big) &(A.4)\cr}$$
     and introducing $S^\prime,\ V^\prime,\ T^\prime,\ {\rm and}\
     W^\prime$ similarly by replacement of $\psi_2 \leftrightarrow
     \psi_4$, the Braden's formula [14] enables us to calculate
     $$\eqalignno{S^\prime &= {1 \over 8}\ (S+V+T+W) &(A.5)\cr
     V^\prime &= {1 \over 8}\ (7S-5V+3T-W) &(A.6)\cr
     T^\prime &= {1 \over 8}\ (21S+9V+T-3W) &(A.7)\cr
     W^\prime &= {1 \over 8}\ (35S-5V-5T+3W) \quad . &(A.5)\cr}$$
     Here, we assumed $\psi_j$'s to commute (rahter than anti-commute)
     with each other.  Especially, we will have two invariants
     $$\eqalignno{&4S^\prime + V^\prime + T^\prime = 4S + V + T \quad ,
     &(A.9)\cr
     &5S^\prime + W^\prime = 5S + W &(A.10)\cr}$$
     as well as antisymmetric combinations of
      $$\eqalignno{&3V^\prime - T^\prime  = -3V  + T
     &(A.11)\cr
     &-7S^\prime + 4V^\prime  + W^\prime = 7S -4V -W \quad . &(A.12)\cr}$$
   \noindent From these together with Eqs. (4.18), it is not hard to derive
   $$\eqalign{<w|&\gamma_\mu x><y|\gamma^\mu z> + <w|y><z|x> -<w|z><x|y>\cr
   &= <w|\gamma_\mu y><z|\gamma^\mu x> + <w|z><x|y> - <w|x><y|z>\cr
   &= <w|\gamma_\mu z><x|\gamma^\mu y> +<w|x> <y|z> -<w|y><z|x>\cr
   &= <w|[x,y,z]>\cr}\eqno(A.13)$$
   by replacing $\psi_1,\ \psi_2,\ \psi_3,\ {\rm and}\ \psi_4$ by $w,\
   x,\ y,\ {\rm and}\ z$.  Eq. (A.13) is equivalent to Eq. (4.17).

     \vfill\eject

     \noindent {\bf \underbar{References}}

     \medskip

     	\item{1.} E.R. Caianiello, \lq\lq On the Universal Fermi-type
     	Interaction I, II, III'', Nuovo Cimento {\bf 8} (1951) 534-541,
     	{\bf 8} (1951) 749-767 and {\bf 10} (1953) 43-53.

     	\item{2.} E.R. Caianiello, \lq\lq Universal Fermi-type
     	Interaction'', Physica {\bf 18} (1952) 1020-1022.

     	\item{3.} E.R. Caianiello, \lq\lq An Argument Against the Majorana
     	Theory of Neutral Particles'', Phys. Rev. {\bf 86} (1952) 564-565.

     	\item{4.} E.R. Caianiello and S. Fubini, \lq\lq On the Algorithm of
     	Dirac Spurs'', Nuovo Cimento {\bf 9} (1952) 1218-1226.

     	\item{5.} R.H. Good, \lq\lq Properties of the Dirac Matrices'',
     	Rev. Mod. Phys. {\bf 27} (1955) 187-211.

     	\item{6.} H. Boerner, \underbar{Representation Theory of Groups}
     	(North Holland, Amsterdam, 1963).

     	\item{7.} L. Jansen and M. Boon, \underbar{Theory of Finite Groups
     	and Applications in Physics} (Wiley, New York, 1967).

     	\item{8.} I. Proteous, \underbar{Topological Geometry} (van Nostand
     	Rheinhold, London, 1969).

     	\item{9.} G.N. Hile and P. Lounesto, \lq\lq Matrix Representations
     	of Clifford Algebras'', Linear Alg. Appl. {\bf 128} (1990) 51-63.

     	\item{10.} e.g. see R.S. Pierce, \underbar{Associative Algebras}
     	(Springer-Verlag, New York, Heidelberg, Berlin 1980).

     	\item{11.} S. Okubo, \lq\lq Real Representations of Finite Clifford
     	Algebras I. Classification, and II. Explicit Construction and
     	Pseudo-octonion'', Jour. Math. Phys. {\bf 32} (1991) 1657-1668,
     	1669-1673.

     	\item{12.} D. Li, C.P. Poole and H.A. Farich, \lq\lq A General
     	Method of Generating and Classifying Clifford Algebras'', Jour.
     	Math. Phys. {\bf 27} (1986) 1173-1180.

     	\item{13.} K.M. Case, \lq\lq Biquadratic Spinor Identities'', Phys.
     	Rev. {\bf 97} (1955) 810-823.

     	\item{14.} H.W. Braden, \lq\lq A New Expression for the
     	$D$-dimensional Fierz Coefficients'', Jour. Phys. {\bf A17} (1984)
     	2927-2934..

     	\item{15.} S. Okubo, \lq\lq Triple Products and Yang-Baxter
     	Equation I. Octonionic and Quaternionic Triple System'', Jour.
     	Math. Phys. {\bf 34} (1993) 3273-3291.

     	\item{16.} R.D. Schafer, \underbar{An Introduction to
     	Non-associative Algebras} (Academic, New York, 1966).

     	\item{17.} B. de Wit and H. Nicolai, \lq\lq The Parallelizing $S^7$
     	Torsion in Gauged $N=8$ Supergravity'', Nucl. Phys. {\bf B231}
     	(1984) 506-532.

     	\item{18.} F. G$\ddot {\rm u}$rsey and C.H. Tze, \lq\lq Octonionic
 Torsion on
     	$S^7$ and Engler's Compactification of $d=11$ Supergravity'', Phys.
     	Lett. {\bf B127} (1983) 191-196.

     	\item{19.} S. Okubo, \underbar{Octonions and Non-associative
     	Algebras in Physics} (Cambridge Univ. Press, Cambridge) to appear.

     	\item{20.} H.J. de Vega and H. Nicolai, \lq\lq Octonionic
     	$S$-matrix'', Phys. Lett. {\bf B244} (1990) 295-298.

        \item{21.} A. Elduque and H.C. Myung, \lq\lq On Flexible
        Composition Algebras'', Comm. Algebra {\bf 21} (1993) 2481-2505.

\end